\documentclass[runningheads]{llncs}

\usepackage[nolist,nohyperlinks]{acronym}
\usepackage{color}
\usepackage{hyperref}
\usepackage{graphicx}
\graphicspath{{figures/}}
\usepackage[labelformat=simple]{subcaption}
\usepackage{xspace}
\usepackage{multirow}
\usepackage{makecell} %
\usepackage[ruled,vlined]{algorithm2e}

\usepackage{ulem}
\newcommand{\eg}{e.g.\xspace}
\newcommand{\wm}{Wikimedia\xspace}
\newcommand{\etal}{et~al.\xspace}
\newcommand{\eui}{\ac{EUI-64}\xspace}
\newcommand{\wifi}{Wi-Fi\xspace}
\newcommand{\vfour}{IPv4\xspace}
\newcommand{\vsix}{IPv6\xspace}
\newcommand{\icmpvsix}{ICMPv6\xspace}
\newcommand{\wiki}{Wikipedia\xspace}

\newcommand{\parhead}[1]{\medskip \noindent \textbf{#1}\hskip .1in}
\newcommand{\Parhead}[1]{\noindent \textbf{#1}\hskip .1in}

\usepackage[override]{cmtt} %

\begin{document}
\title{WikIPedia: Unearthing a 20-Year History\\of IPv6 Client Addressing}
\author{Erik Rye\inst{1,2}\orcidID{0000-0002-8151-8252} \and 
 Dave Levin\inst{2}\orcidID{0000-0003-4957-5131}}
\authorrunning{E. Rye and D. Levin}

\institute{Johns Hopkins University \and
University of Maryland 
}

\maketitle

\begin{abstract}
Due to their article editing policies, \wm sites like \wiki have become
inadvertent time capsules for \vsix addresses. When \wm users make edits
without signing into an account, their IP addresses are used in lieu of a
username. \wm site dumps therefore provide researchers with over two decades
worth of timestamped client \vsix addresses to understand address assignments
and how they have changed over time and space. 

In this work, we extract 19M unique \vsix addresses from \wm sites like \wiki
that were used by editors from 2003 to 2024. We use these addresses to
understand the prevalence of \vsix in countries corresponding to \wm site
languages, how \vsix adoption has grown over time, and the prevalence of EUI-64
addressing on client devices like desktops, laptops, and mobile phones.

\end{abstract}

\section{Introduction}
\label{sec:intro}

In the 1993 film \emph{Jurassic Park}, the eccentric billionaire John Hammond
resurrects long-extinct fauna by extracting infinitesimal amounts of
their DNA from mosquitoes trapped in amber. The DNA in these mosquitoes serves
as a historical blueprint for reverse engineering myriad dinosaur species, which
eventually populate his eponymous theme park.

In the field of network measurement, Wikimedia sites may well be the analog
of mosquito-entrapping amber. When Wikipedians---as editors of the
sites are known---make edits to pages without logging in, their public
IP address is used in lieu of a username. Thus, the sites (which preserve all
historical edits, even those later reversed) and its 24 year history
unintentionally function as an archive for historical Internet data. The IP
addresses logged-out Wikipedians used are, in essence, frozen in amber along
with the timestamp when they were used.

This type of longitudinal data is rare. Few other datasets match the wide
timespan over which it was collected; its earliest entries predate
most social media and many other mainstays of today's web experience. And
datasets that do match its prodigious timeline do not typically include its type
of data.  Routeviews' \ac{BGP} data archives~\cite{rv}, for instance, match the
timespan of Wikimedia's existence. However, \acp{AS} may advertise large swathes
of unused address space, particularly in \vsix, and simply knowing which
prefixes were historically advertised obscures the rich detail present in
understanding real address assignments. 

\vsix research benefits significantly from knowing active \vsix addresses.  For
instance, the \vsix Hitlist~\cite{hitlistwebsite,expanse} regularly provides
researchers with lists of known-active addresses to use as targets of active
measurement campaigns or as training data for
\acp{TGA}~\cite{williams2024seeds,steger2023target,williams20246sense,Murdock:2017:TGI:3131365.3131405,imc18beholder,6loda,shen20256trace,det,6gcvae,6veclm,6scan}.
Similarly, the \vsix Observatory~\cite{rye2023hitlists} releases the /48 prefixes
of NTP clients that visit its servers on a weekly basis.

In this work, we use archival Wikimedia data to extract the client IP addresses
of editors from 2001 through December 2024. We find almost 19 million unique
\vsix addresses, compared to 107 million \vfour addresses, that are used in
lieu of Wikipedia usernames. Our IP address data spans the gamut of sites under
the Wikimedia {\ae}gis: from the popular primary ``encyclopedia'' sites for
dozens of languages, to crowd-sourced textbooks, dictionaries, and quotes, as
well.

In mining this rich vein of historical \vsix addresses, this work makes the
following primary contributions:

\begin{enumerate}
    \item We analyze global \wm \vsix statistics, including i) differences between
        various \wm sites and languages, ii) temporal aspects of \wm \vsix client
        addressing, and iii) changes in \wm \vsix data corresponding with major
        network events.
   \item We examine \eui \vsix addresses in \wm data and discover that this
    obsolescent type of address is actually undergoing a modern revival.
    \item We compare our \wm \vsix corpus with the \vsix Hitlist to understand the
information gained from this unique dataset. 
\end{enumerate}

\section{Background and Related Work}
\label{sec:background}

\subsection{\wm}

The \wm Foundation~\cite{wikimedia} supports a wide variety of free
knowledge projects, including \wiki, Wikibooks, Wiktionary, Wikiquote,
and others, that allow users to freely access, contribute, and verify
its content.  These projects are available in a wide variety of
languages; \wiki, for instance, is available in over 300 languages. 

Most users of \wm sites do not contribute information to these wikis in
the form of new articles or edits to existing ones. However, some users
do; these contributors are encouraged to register an account to which
their edit history will be attached. In early 2025, for instance, the
English Wikipedia reports over 49M registered users with about a third
(15M) having committed at least one edit~\cite{wikipedians}.

\begin{figure*}[t]
\centering
    \includegraphics[width=\textwidth]{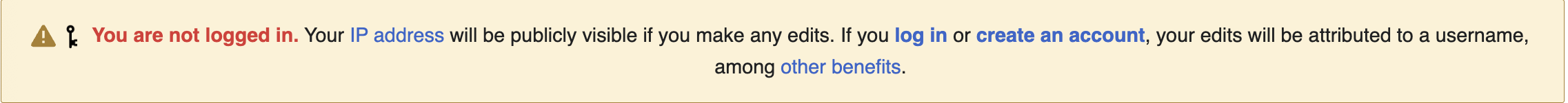}
    \caption{Wikipedia warns logged-out users that their IP address will be
    publicly visible.}
\label{fig:warning}
\end{figure*}

However, \wm sites also allow users to contribute edits without logging
in. When contributors submit edits without logging in, a warning
(Figure~\ref{fig:warning}) alerts them that their IP address will be
used in lieu of a username and will be publicly visible.

\wm periodically publishes ``dumps'' of all of its constituent
sites~\cite{dumps}. Each constituent site and language combination is
available separately, so that consumers of this data can differentiate
between English Wikipedia and Mandarin Chinese Wikiquote entries, for
instance. Importantly, these archives contain both current and
\emph{historical} data, including edits that were made long ago and
have since been overwritten, as well as edits that were later reverted.

\subsection{Related Work}
\label{sec:related}

\parhead{Studies of \wm}
This work uses \wm sites as a source of active client \vsix addresses over time.
It builds on several studies that measure \wiki generally, and make use of IP
addresses found in \wiki edits specifically.

A significant body of work has studied \wiki content.
Voss authored an early measurement of \wiki, counting the number of
articles, authors, and edits across multiple language versions of the
site~\cite{voss2005measuring}. 
Unlike these prior studies, our work does not focus on the wikis'
\emph{content}, but rather the IP addresses of those who help curate
it.

Wikipedians' IP addresses have been studied in the context of Wikipedia
vandalism---that is, removing accurate information from articles, removing
articles entirely, inserting inaccurate information into articles, or otherwise
making ``unproductive'' edits~\cite{west2010detecting,kiesel2017spatio}. Others have used \wm address
edit history to filter for Tor exit nodes~\cite{tran2020anonymity}. By contrast,
we are interested in \emph{all} edits recorded by \vsix addresses in the \wm
dumps.

Almeida~\etal performed a broad study of how user behavior on \wiki had
evolved from 2001 to 2006~\cite{almeida2007evolution}.
As part of this larger study, they reported finding 3.8M unique IP
addresses in \wiki edits over this nearly six-year span.
By contrast, our work spans over two decades, and examines primarily
\vsix addresses, which were not widely logged by \wm sites until after
2012 (\S\ref{sec:results}).

Finally, Zander~\etal extracted \vfour addresses from \wiki edit logs
over the three-year period from 2011 to
2014~\cite{zander2014capturing}. They used the \vfour addresses they
obtained to improve insights into \vfour address space exhaustion
beyond the visibility achieved through active measurement campaigns.
We focus exclusively on \vsix in this work, and our study spans over
two decades (2003--2024).

\parhead{Gathering \vsix addresses}
This work seeks to gather a large corpus of \vsix addresses, which has
been the focus of many recent
studies~\cite{hitlistwebsite,expanse,williams2024seeds,steger2023target,williams20246sense,Murdock:2017:TGI:3131365.3131405,imc18beholder,rye2023hitlists}.
These rely on a range of passive and active measurements, and have
resulted in datasets orders of magnitude larger than we are able to
obtain by looking only at \wm sites.
However, these prior efforts seek to obtain \vsix addresses that are
live and in active use \emph{right now} (primarily to guide future scanning
efforts); in contrast, our work provides a \emph{historical} view of
\vsix addresses.

\section{Data Collection Methodology}
\label{sec:methodology}

\Parhead{\wiki historical data}
We obtain \vsix addresses from edits to \wiki and other Wikimedia sites.
\wm publishes ``dumps'' of their sites' content: every article's text, links to
every picture, and---critically---all edit metadata going back to the sites'
inception in 2001.
We obtained the December 2024 dumps for all available content sites.

The most common site type from the December 2024 dumps is Wikipedia, the
flagship encyclopedia site. There are 394 individual Wikipedia sites that span
both languages (\eg, enwiki and dewiki, for the English and German versions of
Wikipedia, respectively\footnote{Wikimedia uses ISO-639 2-letter language
codes as prefixes and the site type as suffixes for the full \wiki site
identifier.}) and special events (\eg, Wikimania, Wikipedia's annual conference,
had its own public wiki from 2005--2018).  Wikitionaries, which are
user-editable dictionaries, are the second-most common site type, with 195
individual Wiktionary sites spanning many languages.  Table~\ref{tab:wikitypes}
in the Appendix lists the number of individual sites per category.

\parhead{Extracting IP addresses}
After downloading this corpus from Wikimedia, we parse the historical edit data
to obtain edits that were authored by an unregistered user.
Because Wikimedia's policy is to log and use the IP address of unregistered
users, this is equivalent to filtering edits for those authored by an IP address.
This process is highly unlikely to result in false positives, as \wiki
specifically prohibits usernames that are (or even look like) an IP
address~\cite{usernamepolicy}. 
We separate these by IP version. In this work, we are primarily interested in
\vsix addresses due to the challenges of obtaining large-scale client \vsix
addresses, but retain \vfour addresses for comparison purposes in
\S\ref{sec:results}.

\parhead{Contemporaneous AS data}
Finally, we also used Routeviews' \vsix BGP RIB dumps~\cite{rv}. In order to
determine what \ac{AS} a historical IP address was in at the time that it was
logged, we need to be able to look up these IP addresses in BGP data
contemporaneous to their appearance in a \wm site. After obtaining the
Routeviews data, we looked each \vsix address in our dataset up in the
chronologically closest Routeviews \vsix RIB dump to obtain an ASN. As
Routeviews produces BGP dumps every two hours, the time delta between an IP
address's collection and its AS BGP lookup is rarely more than an hour.

\parhead{Limitations}
Our \wm dataset offers a unique view into the history of client IPv6
addresses, but it is not without limitations.
First, it is limited to client devices; the IPv6 addresses we extract from the
\wm dumps are associated with the user who was submitting the edits to the
wiki.
It is therefore highly unlikely that our data comprises addresses of web servers,
routers, and so on. Further, relatively few users make edits to \wm pages, and
even fewer make edits without logging into an account. 
Second, while it has broad reach, \wiki is not ubiquitous; several countries
block access to \wiki (notably, China). This naturally creates bias in our
dataset toward regions where \wiki is more accessible and popular, as
populations without access or desire to visit \wm sites will do less editing.  
Finally, it is difficult to reason about how representative the dumps' IP
addresses are of the entirety of the IPv6 space. 
However, it is encouraging that the \wm dumps contain hundreds of
different language-specific sites, as they likely capture users from the regions that
speak those languages.
\section{Results}
\label{sec:results}

In this section, we analyze the \vsix client addresses we obtained from the
dumps of 1,005 \wm sites, and compare this dataset to other contemporaneous
data. 

\begin{figure}[t]
\centering
    \includegraphics[width=0.65\linewidth]{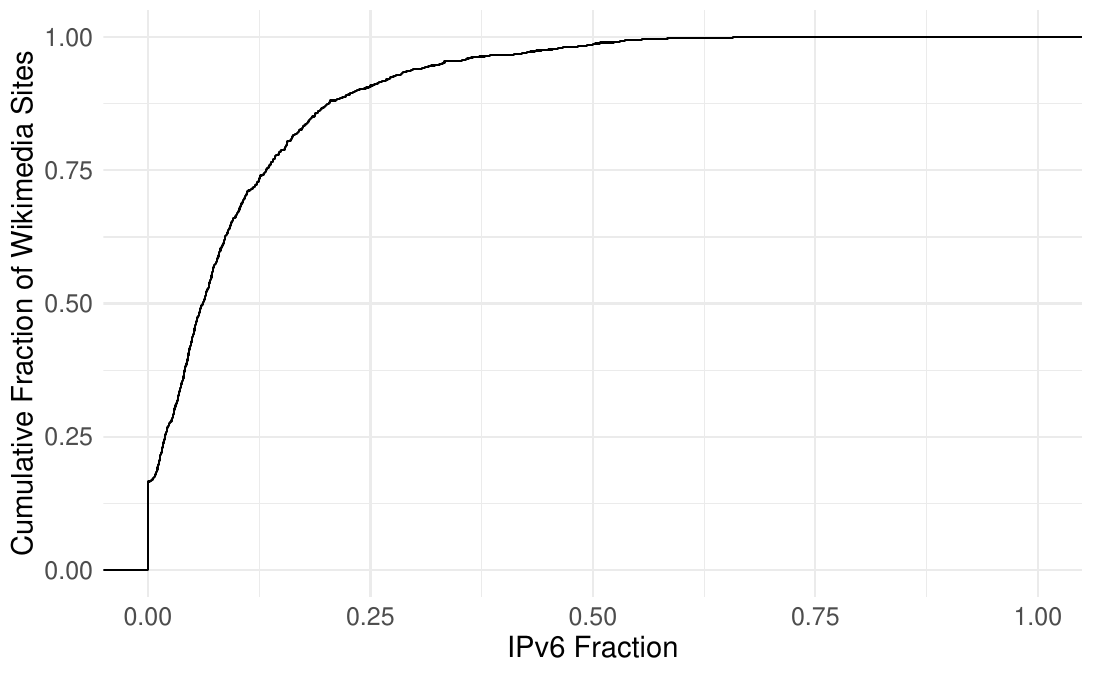}
    \caption{Fraction of \vsix addresses of the total number of IP addresses
    logged per \wm site.}
\label{fig:v6-fraction}
\end{figure}

\subsection{\vsix Address Frequency}

In parsing the \wm site data dumps and extracting \vsix addresses from
them, we obtained 19,292,487 unique \vsix addresses. This is
approximately 0.18$\times$ the number of unique \vfour addresses used
as user identifiers over the same period (107,371,338).
The fraction of \vsix addresses of all IP addresses logged per \wm site
varies between 0 and 0.66, though the median value is 0.06.
Figure~\ref{fig:v6-fraction} shows the fraction of \vsix addresses
as a CDF of \wm site.

Interestingly, because some \wm sites are highly regional, their
proportion of \vsix addresses indicates regional adoption rates of
\vsix.
In the tail of the distribution are several sites with only a small number of
total addresses, such as the Assamese Wikiquote site, which has the highest
\vsix fraction of 0.66 (12 \vfour and 23 \vsix addresses).
However, the Hindi
\wiki site contains both a large number of total unique IP addresses, with
297,741, and a large fraction of \vsix addresses, with 0.57 (168,703). 
The Serbian \wiki has an approximately equal number of total IP addresses
logged, with 267,377. Conversely, however, it has an extremely low fraction of
\vsix addresses to total IP addresses, at 0.03 (259,118 \vfour to 8,259).  

The significant variation in the proportion of \vsix addresses across
different \wm sites is likely due to the adoption rate of \vsix in the
countries the language-specific wikis are spoken in.
RIPE reports an \vsix adoption rate of 77\% in India, where Hindi is
primarily spoken, compared with an \vsix adoption rate of 7\% in
Serbia, where Serbian is primarily spoken~\cite{ripe-adoption}.

\begin{figure}[t]
\centering
    \includegraphics[width=0.75\linewidth]{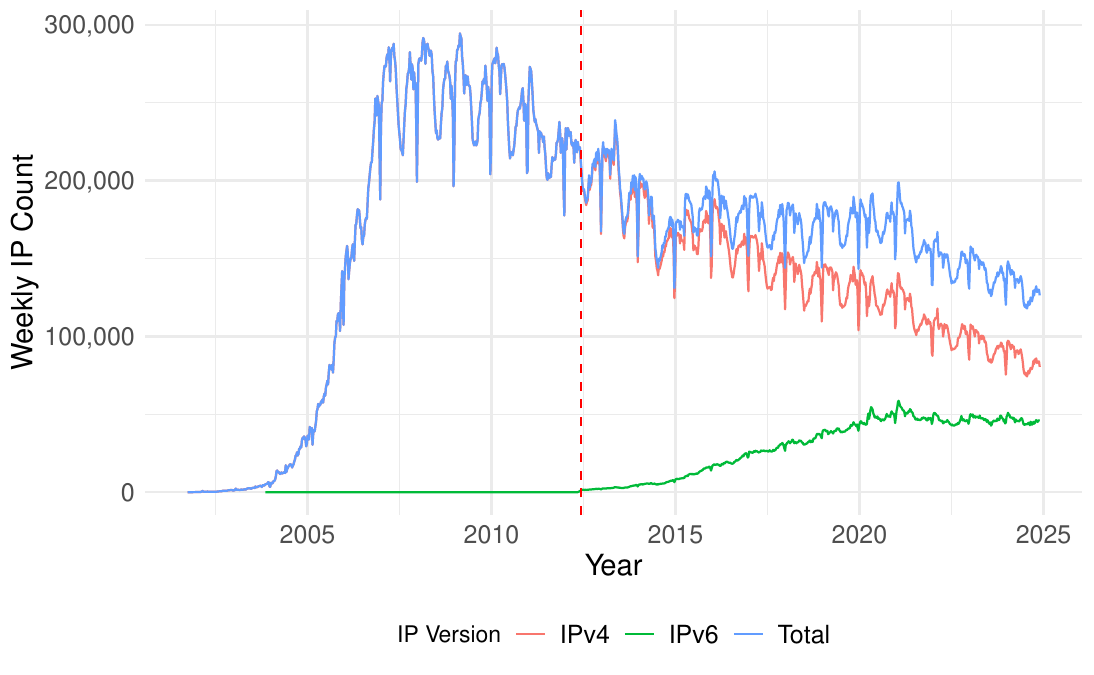}
	\caption{Unique IP addresses per week by protocol version. Red
	dashed line is World IPv6 Launch day.}
\label{fig:addresses-by-week}
\end{figure}

Not all \wm sites have logged \vsix addresses. Of the 1,005 \wm for
which we have data, only 828 (82\%) have at least one \vsix address
logged. There is at least one \vfour address logged as a user
identifier for 993 (99\%) \wm sites.
The English Wikipedia site contributes half of the total number of
unique logged \vsix addresses; the German, French, Japanese, and
Spanish Wikipedia sites contribute more than 5\%.
Table~\ref{tab:site-counts} in the Appendix lists the top \wm sites.

\subsection{Temporal Characteristics}

\begin{figure*}[t]
    \begin{subfigure}[c]{0.5\textwidth}
        \centering
        \includegraphics[width=\linewidth]{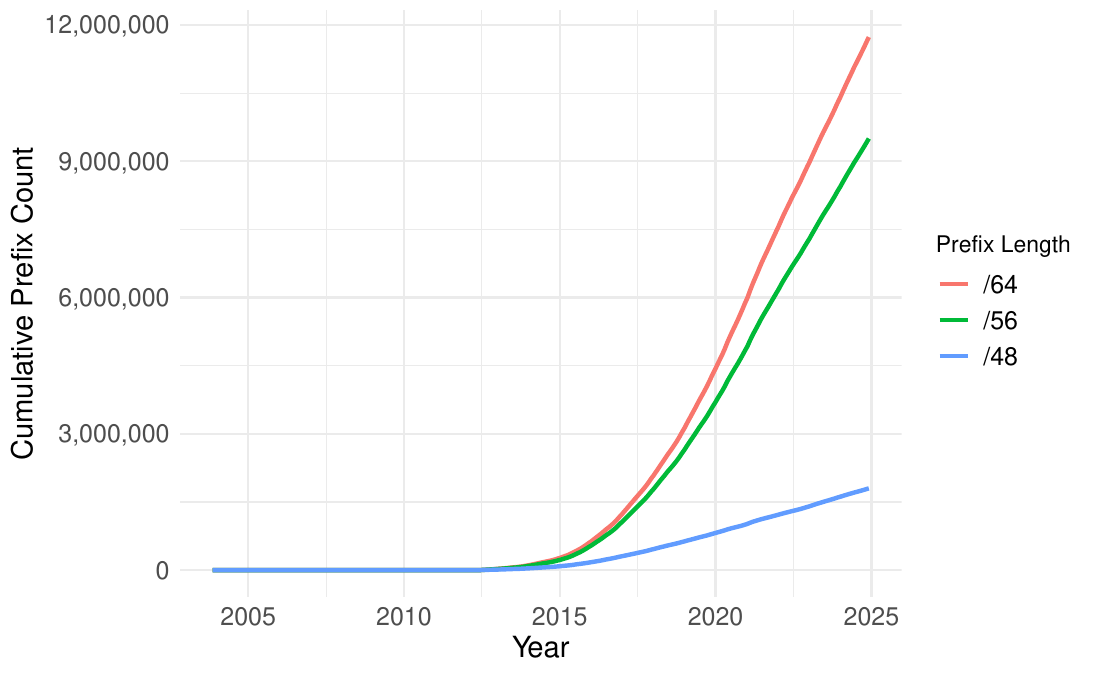}
        \caption{Cumulative numbers of prefixes observed in the \wm corpus
        ($N=19,292,487$ /128s).}
\label{fig:cumu-prefix}
  \end{subfigure}
    \hspace{1em}
    \begin{subfigure}[c]{0.5\textwidth}
        \centering
        \includegraphics[width=\linewidth]{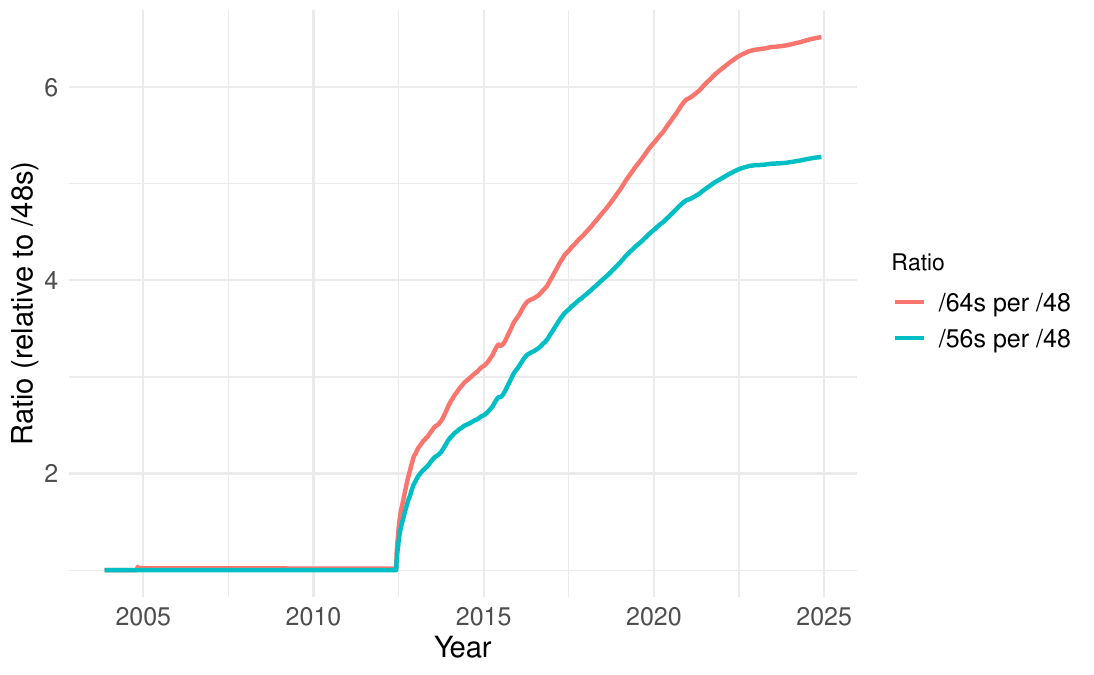}
        \caption{/56 and /64 ratio per /48 over time.}
        \label{fig:ratio}
  \end{subfigure}
    \caption{Prefix observations in the \wm corpus.}
    \label{fig:prefixes}
\end{figure*}

\begin{figure}[t]
\centering
    \includegraphics[width=0.75\linewidth]{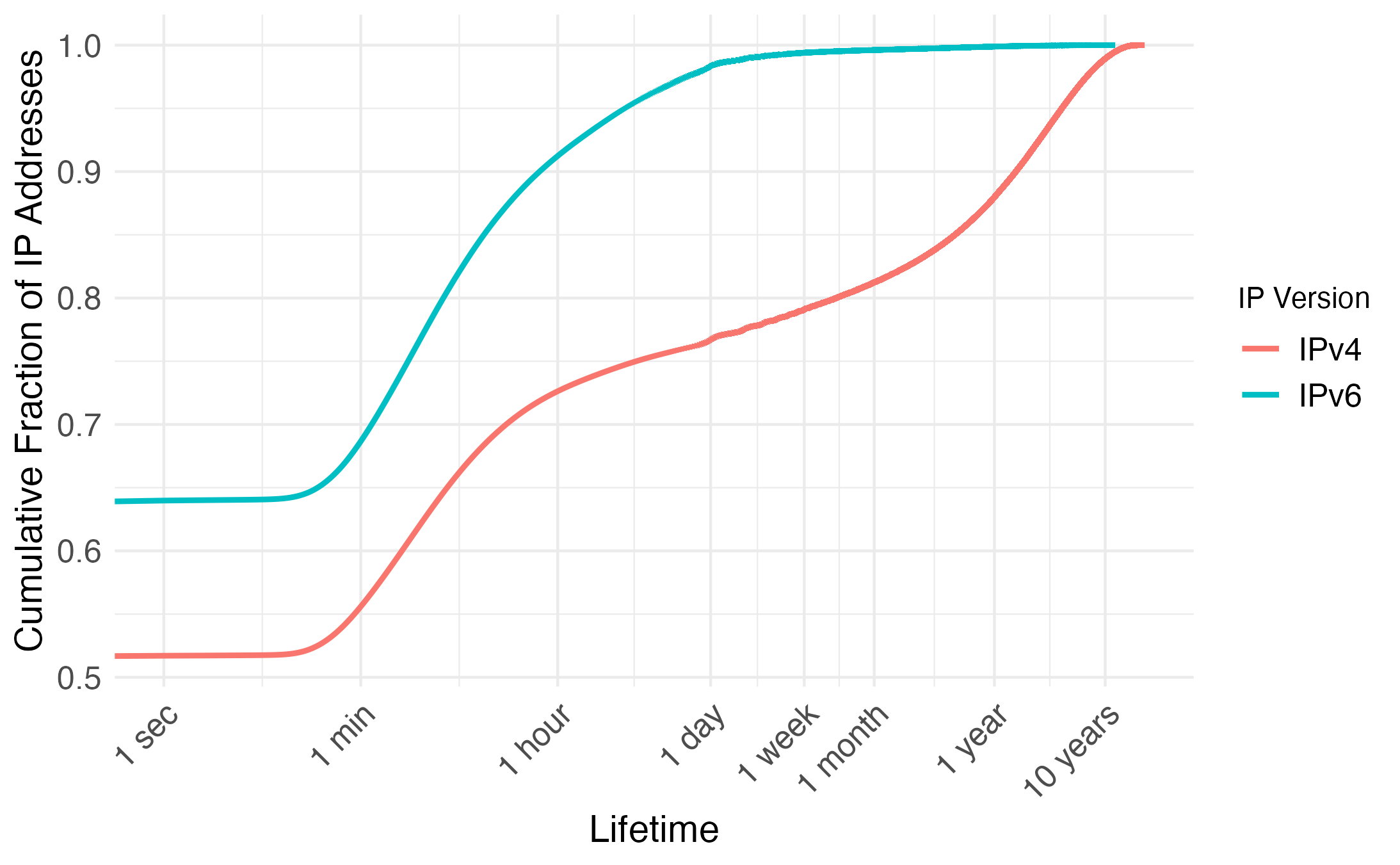}
    \caption{Lifetimes of the IP addresses logged in the \wm dataset by IP
    protocol version.}
\label{fig:ip-version-lifetime}
\end{figure}

\vsix addresses used as edit identifiers in \wm sites appear between November
2003 to December 2024, the month that we obtained the \wm data. However,
\vsix addresses as edit identifiers remained relatively sparse from \wm's
inception through the mid-2010s. 

Figure~\ref{fig:addresses-by-week} displays the number of unique weekly IP
addresses logged by the 1,005 \wm sites since 2001 across both versions of IP.
While \vsix addresses first appear in 2003, \vfour addresses are first logged in
2001 and dominate throughout the first decade of \wm's existence. \vsix
addresses start to increase shortly after World IPv6 Launch day, which occurred
on 6 June 2012 and \wm participated in~\cite{world}, and is annotated in Figure~\ref{fig:addresses-by-week} as a
red, vertical dashed line. At the end of 2024, roughly twice as many \vfour
addresses appeared as \vsix addresses.

Figure~\ref{fig:cumu-prefix} depicts the number of cumulative /48, /56, and /64
prefixes observed in the \wm \vsix corpus over time. While best practice
recommends /48 and /56 as prefix delegation sizes to customer end sites, in
practice, ISPs may assign /48, /52, /56, /60 and /64~\cite{bcop-prefix}. This
helps us estimate the number of unique clients with \vsix addresses logged in
the \wm corpus. While the number of total /48s is relatively low ($\sim$1.8M),
the number of /64s is approximately 60\% of the total number of unique /128s.
This indicates that 40\% of \wm \vsix addresses come from the same /64 networks,
which strongly suggests edits were made by users that are part of the same home or
campus network, if not the same individuals themselves. Figure~\ref{fig:ratio}
shows that the number of /56s and /64s per /48 increased rapidly through 2022.
However, the number of observed subnets per /48 has flattened from 2023 onward,
indicating relatively stable number of observed /56 and /64 subnets per /48.

Finally, Figure~\ref{fig:ip-version-lifetime} displays the length of time
between the first and last observations of each IP address in the \wm dataset.
Most addresses ($\sim$52\% of \vfour and $\sim$64\% of \vsix) are only
observed once in the \wm dump, which manifests as a lifetime of 0. Because best
practice recommends that \vsix client addresses be both random and
ephemeral~\cite{rfc4941}, \vsix address lifetimes are significantly shorter than
\vfour address lifetimes. Further, \vfour addresses may have \emph{many}
clients behind a device performing \ac{NAT}, whereas \ac{NAT} is extraordinarily
rare in \vsix, and most \vsix addresses correspond to a single host.

\subsection{AS Contributions}

\begin{figure}[t]
\centering
    \includegraphics[width=0.75\linewidth]{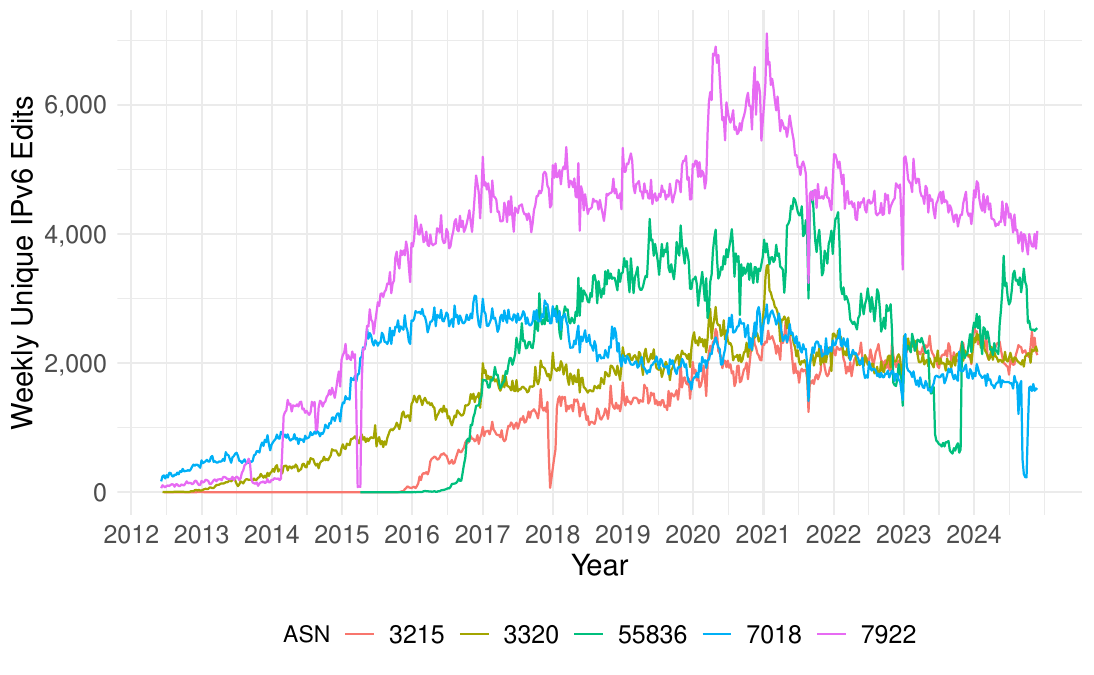}
    \caption{Number of unique \vsix addresses per week for the top 5 \acp{AS}.}
\label{fig:addresses-by-week-by-as}
\end{figure}

\begin{figure*}[t]
    \begin{subfigure}[c]{0.5\textwidth}
        \centering
            \includegraphics[width=\linewidth]{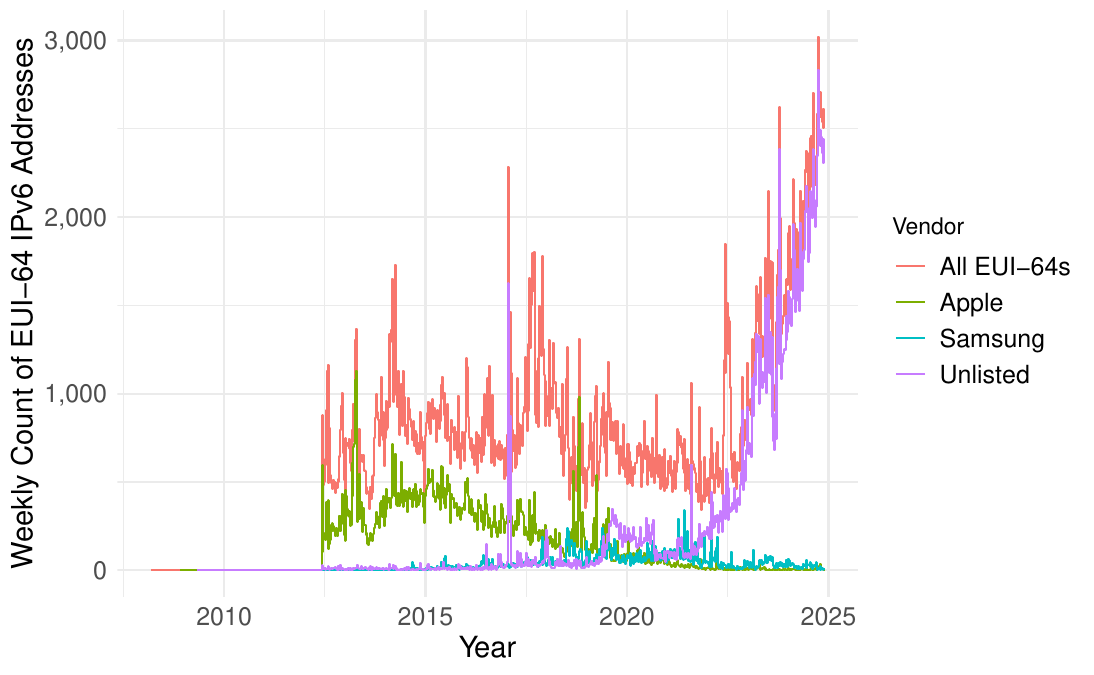}
            \caption{Number of \eui \vsix addresses per week.}
            \label{fig:eui64-by-week}
   \end{subfigure}
     \hspace{1em}
     \begin{subfigure}[c]{0.5\textwidth}
         \centering
            \includegraphics[width=\linewidth]{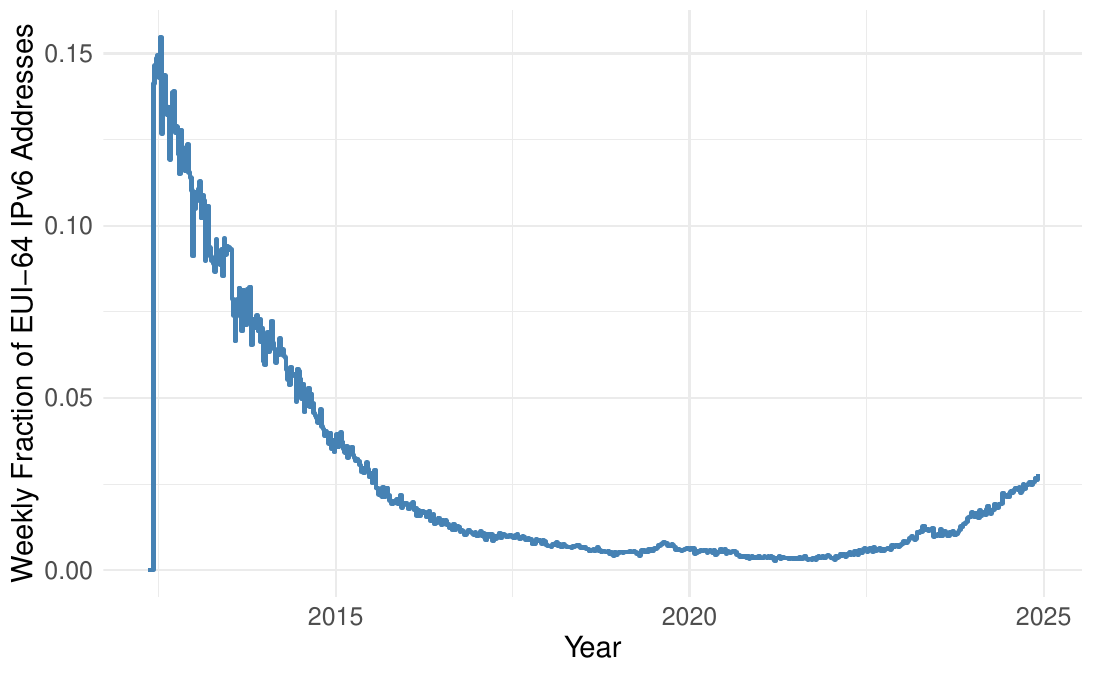}
            \caption{The fraction of \eui \vsix addresses of all \vsix addresses per
    week.}
            \label{fig:eui64-fraction}
   \end{subfigure}
     \caption{EUI-64 addresses in the \wm corpus.}
     \label{fig:eui-combined}
 \end{figure*}

Next, we examine the number of \vsix addresses from \wm edits according to their
\ac{AS}. Because the \ac{AS} to which an \vsix address belongs
may change over time as address blocks are reallocated and reassigned,
we use contemporaneous \ac{BGP} RIB data from Routeviews to look up the \ac{AS} of an
IP address in the chronologically-closest RIB dump.

Figure~\ref{fig:addresses-by-week-by-as} displays the number of unique weekly
\vsix addresses per week observed in the \wm site dumps for each of the top five
\acp{AS}. While Comcast (AS7922), AT\&T (AS7018), and Deutsche Telekom (AS3320)
have had \vsix addresses in the \wm corpus since 2012, other \acp{AS}
did not deploy \vsix until later. This is reflected in
Figure~\ref{fig:addresses-by-week-by-as}. For instance, Reliance Jio (AS55836),
an Indian telecom provider, did not deploy \vsix widely until September
2016~\cite{reliance-blog}. This deployment date tracks with AS55836's curve in
Figure~\ref{fig:addresses-by-week-by-as}, and correlates with contemporaneous
\vsix availability measurements, \eg from APNIC~\cite{apnicjio}. 

Finally, note that the Figure~\ref{fig:addresses-by-week-by-as} displays sharp
decreases for many of the \acp{AS}. For instance, Comcast (AS7922) exhibited a
large decrease in \vsix addresses logged in early 2015, while Reliance Jio
(AS55836) had an extended dip in \vsix edits over a three-month period in 2023.
While we were unable to definitively link these to publicized network events, we
speculate that these decreases in \vsix edits occurred due to changes in \vsix
address assignment policies or routing by these networks. 

\subsection{\eui Addresses}

Next, we turn to an analysis of the devices manufacturers of the logged \wm
\vsix addresses. To do this, we filter for the subset of \vsix addresses that
are \eui. \eui addresses embed the \ac{MAC} address of the
device's interface into the lower 64 bits of the \vsix address\footnote{The MAC
address is extended to 64 bits by inserting the bytes \texttt{0xfffe} between
its third and fourth bytes; the second-least significant bit of the
most significant byte of the MAC address (the Universal/Local (U/L) bit) is
typically also inverted.};
because \ac{MAC} addresses frequently encode the manufacturer of the device in
the MAC address's
upper three bytes, we can determine the type of devices unregistered \wm users
were using for this subset.

Of the 19M total \vsix addresses logged, 167,417 (0.87\%) of are \eui.  These
\eui \vsix addresses contain 145,832 unique \ac{MAC} addresses.
Table~\ref{tab:manufs} in the Appendix lists the number of unique \ac{MAC} addresses by the
manufacturer derived from looking up the upper three bytes (the \ac{OUI}) of
each MAC address in the public list of IEEE \ac{OUI} assignments~\cite{oui}.
While prior work has found large numbers of \eui addresses assigned to
routers~\cite{rye2021follow} and IoT
devices~\cite{saidi2022one,rye2023hitlists}, \eui addresses in the \wm dataset
largely belong to mobile, laptop, and desktop clients. For instance, Apple is
the most commonly resolved \ac{OUI} vendor, along with HP, Intel, ASUS, and Samsung.
However, more than half (56\%) of the EUI-64-derived MAC addresses did not resolve to
an IEEE-assigned OUI, a phenomenon also observed by other recent work~\cite{rye2023hitlists}.
This suggests that some devices may be using \eui \vsix addresses in conjunction
with randomized \ac{MAC} addresses, a common privacy protection employed by
mobile
devices~\cite{vanhoef2016mac,matte2016defeating,martin2017study,fenske2021three,uras2022mac}.

\eui addresses are well-known to present a privacy risk to users, as they allow
long-term tracking of a static identifier~\cite{rfc4941}. Surprisingly, the
number of edits coming from \eui addresses has \emph{increased} over time.
Figure~\ref{fig:eui64-by-week} depicts the number of \eui\vsix addresses making
edits per week. This is due in large part to the ``Unlisted'' MAC addresses,
whose \acp{OUI} are not in the IEEE OUI database. 
Figure~\ref{fig:eui64-by-week} also shows that while Apple historically employed \eui
addressing in the mid-2010s, it has largely phased out these types of addresses.
However, the prevalence of ``Unlisted'' \ac{MAC} addresses has dramatically
increased since 2021. The privacy ramifications of using random MAC addresses to
form \eui \vsix addresses are less severe than embedding a device's true
hardware MAC address, as random MAC addresses obscure the device manufacturer.
Nonetheless, if the random MAC address used in the \eui \vsix address is
sufficiently long-lived (\eg, Android and iOS use a stable, random MAC address
on a per-network basis for most \wifi networks\cite{applerandomaddress,androidrandomaddress}) the longitudinal tracking threat of \eui
\vsix addresses remains.

Not only have the raw number of \eui \vsix addresses increased over time, but
the \emph{fraction} of \vsix addresses that are \eui has simultaneously
increased. Figure~\ref{fig:eui64-fraction} shows that while \eui \vsix addresses
represented less than 1\% of the total number of \vsix addresses seen weekly
between 2017 and 2023 (down from a high of $\sim$15\%), they have become
increasingly prevalent once again, comprising nearly 3\% of \vsix addresses in
December 2024.

\begin{figure}[t]
\centering
    \includegraphics[width=0.75\linewidth]{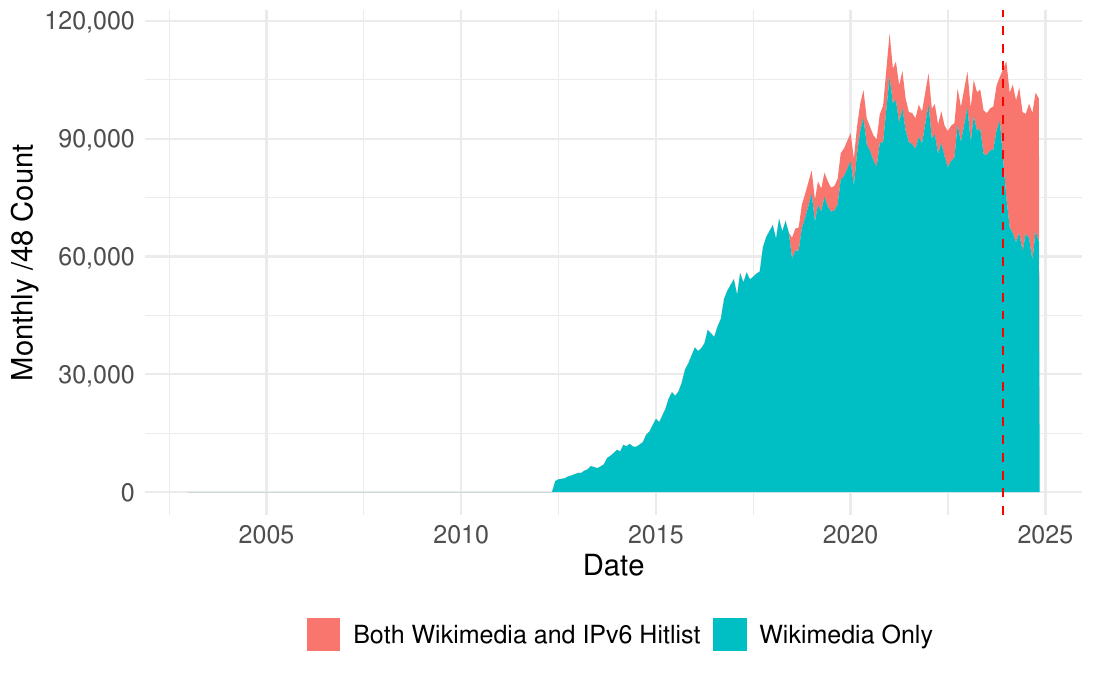}
    \caption{Number of unique /48s in the \wm dumps and overlap with the \vsix
    Hitlist. The red dashed line at December 2023 is when the IPv6 Hitlist began
    incorporating data from IPinfo.}
\label{fig:hitlist-compare}
\end{figure}

\subsection{Comparison with the \vsix Hitlist}

Finally, we compare the \vsix addresses in the \wm corpus to the \vsix
Hitlist's~\cite{hitlistwebsite,expanse,gasserscanning}
\icmpvsix-responsive addresses over the lifetime of the Hitlist.
Figure~\ref{fig:hitlist-compare} displays the number of unique /48 prefixes
contained in the \wm dumps, as well as overlapping /48s with the \vsix Hitlist
(\vsix Hitlist-only /48s omitted for clarity) binned by month. The number of
/48s seen monthly in \wm data reaches $\sim$100k by about 2020. The overlap with
the \vsix Hitlist remains relatively low ($\sim$5-10\%) until December 2023,
when the \vsix Hitlist began incorporating addresses from IPinfo~\cite{ipinfo}.
From this point forward, the amount of overlap with the \vsix Hitlist is
approximately one-third, indicating that the /48s coming from IPinfo are likely
from customer networks.

This demonstrates that the \wm \vsix corpus provides a unique source of \vsix
address data that is difficult to obtain from either active measurements or
without the broad reach than \wm provides. For applications that rely on
obtaining active \vsix addresses (\eg, \vsix TGAs), the \wm corpus provides
another data source currently missing from state-of-the-art hitlists.
\section{Conclusion}
\label{sec:conclusion}

Currently, the network measurement community is focused on collecting as many
active, in-use IPv6 addresses as
possible~\cite{hitlistwebsite,imc18beholder,rye2023hitlists}, so as to enable
the future of Internet-wide scanning.
In this paper, we have instead opted to peer into the past.
By using IPv6 addresses encased in the ``amber'' of decades of edits to \wiki
and other \wm websites, we showed it is possible to reason about significant
changes in IPv6 adoption and policy over decades.

Toward that end, we extracted 19M unique \vsix addresses over 2003--2024,
spanning multiple languages, and ASes around the world.  Our historical view of
\vsix shows how critical the World IPv6 Launch Day truly was for \vsix adoption:
prior to it, \vsix was an almost nonexistent novelty among client devices.
We find that \vsix addresses tend to have shorter lifetimes than their \vfour
peers. The majority ($\sim$64\%) of the \vsix addresses that are logged in \wm
edits appear only once. This is likely an artifact of the \vsix client
addressing best practice to choose client addresses randomly and to change them
periodically. 

We observe \vsix roll-outs occurring by \acp{AS}. When \vsix service is deployed
to customers by ASes, steep upticks in logged \vsix addresses from those
\acp{AS} occur. This allows us to retroactively detect when \vsix is
\emph{deployed}, rather than simply \emph{announced} by the AS using BGP.

Finally, our analyses showed how \eui \vsix addresses---largely considered a
privacy violation~\cite{rye2023hitlists}---had been getting phased out, but have
recently started to see a resurgence. This points to randomized MAC addresses
being used to build \eui addresses, and confirms trends detected in other recent
work.

\begin{credits}
\subsubsection{\ackname} This work was supported by NSF grant 
CNS-2323193.

\end{credits}

\bibliographystyle{splncs04}
\bibliography{conferences,refs}
\appendix

\section*{Appendix A: Ethical Considerations}
\label{sec:ethics}

Users that wish to make an edit to a \wm site page, but do not wish to create
or log into an account, are warned with text that reads ``You are not logged in.
Your IP address will be publicly visible if you make any edits'' (see
Figure~\ref{fig:warning}). Continuing to submit an edit constitutes the consent
of the user. Consistent with this statement, Wikimedia site dumps that contain
user addresses are publicly available~\cite{wikimedia}.

Due to these facts, the collection of the IP addresses logged by Wikimedia sites
does not itself raise ethical issues. As of July 2025, Wikimedia is phasing out
the collection of IP addresses from logged-out users due to GDPR and other
privacy concerns~\cite{policyupdate}; as such, we will not release the set of IP
addresses we extracted from these dumps separately.  

\section*{Appendix B: \wm Sites}

\begin{table}[ht]
    \centering
\caption{Number of Wikimedia sites by wiki type}
\label{tab:wikitypes}
\begin{tabular}{cr||cr}
\textbf{Wiki Type} & \multicolumn{1}{c||}{\textbf{Site Count}} & \multicolumn{1}{c}{\textbf{Wiki Type}} & \multicolumn{1}{c}{\textbf{Site Count}} \\
\hline
Wikipedia          & 394                                     & Wikimedia          & 37                                      \\
Wiktionary & 195                                            & Wikinews                               & 36                                      \\
Wikibooks          & 121                                     & Wikivoyage & 27                                      \\
Wikiquote          & 97                                      & Wikiversity & 18                                      \\
    Wikisource         & 80                                      &
    \textbf{Total} & \textbf{1,005}
\end{tabular}
\end{table}

Table~\ref{tab:wikitypes} lists the number of \wm sites by category; within each
category, most of the individual sites are language-specific variations of the
site type, although others are for special events or uses (\eg, wikis for the
\wiki ``Wikimania'' event.) 

\section*{Appendix C: \vsix Addresses by \wm Site}

\begin{table}[t]
    \centering
    \caption{Number of unique logged \vsix addresses per \wm site; some
    addresses appear in the logs of multiple sites.}
    \label{tab:site-counts}
\begin{tabular}{c|r|c}
\textbf{Wiki Site} & \multicolumn{1}{c|}{\textbf{\begin{tabular}[c]{@{}c@{}}\#IPv6\\ Addresses\end{tabular}}} & \textbf{\begin{tabular}[c]{@{}c@{}}\% of\\ All IPv6\end{tabular}} \\
\hline
English Wikipedia  & 9,638,421                                                                               & 50                                                                \\
German Wikipedia   & 1,518,286                                                                               & 7.9                                                               \\
French  Wikipedia  & 1,457,954                                                                               & 7.6                                                               \\      
Japanese Wikipedia & 1,173,841                                                                               & 6.1                                                               \\
Spanish Wikipedia  & 1,061,633                                                                               & 5.5                                                               \\      
1,000 other        & 5,123,867                                                                               & 26.6                                                              \\      
\hline
\textbf{Total}     & \textbf{19,292,487}                                                                     & \textbf{100}
\end{tabular}
\end{table}

Table~\ref{tab:site-counts} lists the top \wm sites by number of unique \vsix
addresses logged in their site's dump.

\section*{Appendix D: \eui \vsix Addresses}

Table~\ref{tab:manufs} lists the number of distinct MAC addresses observed in
\eui \vsix addresses in the \wm data. We use the IEEE \ac{OUI}
database~\cite{oui} to resolve the MAC addresses to manufacturers.
Surprisingly, the most commonly observed manufacturer is ``unlisted'', meaning
that the \acp{OUI} did not resolve to any manufacturer. 

\begin{table}[t]
    \centering
    \caption{Number of distinct devices per manufacturer as determined from
    \eui \vsix address-embedded MAC addresses.}
    \label{tab:manufs}
\begin{tabular}{c|r||c|r}
\textbf{Manufacturer} & \multicolumn{1}{c||}{\textbf{Count}} & \textbf{Manufacturer} & \multicolumn{1}{c}{\textbf{Count}} \\
\hline
Unlisted              & 82,244                             & ASUSTek               & 2,962                              \\
Apple                 & 23,168                             & HP                    & 1,662                              \\
Samsung               & 5,367                              & Hon Hai Precision     & 1,246                              \\
Intel                 & 4,393                              & 1,141 other           & 21,631                             \\
Dell                  & 3,159                              & \textbf{Total}        & \textbf{145,832}
\end{tabular}
\end{table}
 
\begin{acronym}
  \acro{AS}{Autonomous System}
  \acrodefplural{AS}[ASes]{Autonomous Systems}
  \acro{ASN}{\ac{AS} Number}
  \acro{BGP}{Border Gateway Protocol}
  \acro{CDN}{Content Distribution Network}
  \acro{CPE}{Customer Premises Equipment}
  \acro{DAD}{Duplicate Address Detection}
  \acro{EUI-64}{Extended Unique Identifier - 64}
  \acro{ISP}{Internet Service Provider}
  \acro{IID}{Interface Identifier}
  \acro{LAN}{Local Area Network}
  \acro{NIC}{Network Interface Card}
  \acro{NAT}{Network Address Translation}
  \acro{NTP}{Network Time Protocol}
  \acro{MAC}{Media Access Control}
  \acro{OS}{Operating System}
  \acro{OUI}{Organizationally Unique Identifier}
  \acro{SLAAC}{Stateless Address Autoconfiguration}
  \acro{SOHO}{Small Office-Home Office}
  \acro{TGA}{Target Generation Algorithm}
  \acro{ULA}{Unique Local Address}
  \acro{U/L}{Universal/Local}
  \acro{ULA}{Unique Local Address}
  \acro{VPS}{Virtual Private Server}
  \acrodefplural{VPS}[VPSes]{Virtual Private Servers}
  \acro{WAN}{Wide Area Network}
\end{acronym}

\end{document}